# Bayesian Thought in Early Modern Detective Stories: Monsieur Lecoq, C. Auguste Dupin and Sherlock Holmes

**Joseph B. Kadane**

*Abstract.* This paper reviews the maxims used by three early modern fictional detectives: Monsieur Lecoq, C. Auguste Dupin and Sherlock Holmes. It find similarities between these maxims and Bayesian thought. Poe's Dupin uses ideas very similar to Bayesian game theory. Sherlock Holmes' statements also show thought patterns justifiable in Bayesian terms.

*Key words and phrases:* Arthur Conan Doyle, Edgar Allan Poe, Emile Gaboriau, odd and even, Bayesian Game Theory.

## 1. INTRODUCTION

The three writers considered here, Emile Gaboriau (1832–1873), Edgar Allan Poe (1809–1849) and Sir Arthur Conan Doyle (1859–1930) are considered to be the founders of the modern interest in detective stories, writing even before the term "detective" was used (Bleiler, 1975). In addition to the many novels and short stories that have ensued, there are also popular television crime, mystery and police shows that can be considered progeny. This popularity continues despite the weaknesses in the underlying forensic science, as emphasized by a recent report of the National Research Council (2009).

This essay aims to examine the pattern of thought used by their respective detectives: Monsieur Lecoq, C. Auguste Dupin and Sherlock Holmes. What does it mean to understand the thoughts of a fictional character? With a real person, one can ask questions and run experiments to ascertain how they think. Although indirect methods such as functional magnetic resonance imaging (fMRI) cannot yet be used to tell what a person is thinking, it is not impossible that in the future this may be possible. We have so-called lie-detector machines, although another National Research Council Report (2003) has seriously challenged their accuracy.

On the other hand, there is a sense in which understanding fictional characters is easier than understanding real ones. There is a fixed body of written work, and this is all the evidence there will ever be. Those words tell what characteristics of the detectives the author considers most important. When the author writes about the way such characters go about their work, this can be taken to be authoritative. Thus there is no issue, for example, of changing or molding the responses that a real person might give in response to a question framed in a way the respondent hadn't considered previously. Thus we may take the statements written by the authors as summaries of what they intend their characters (here, detectives) to teach the readers.

## 2. EMILE GABORIAU'S MONSIEUR LECOQ

The plot of the novel *Monsieur Lecoq* (1869) and its sequel *In Honor of the Name* Gaboriau (1975) revolves around the efforts of novice detective Lecoq to establish the identity of a prisoner who killed three people in a brawl in a bar on the outskirts of Paris. The intricacies are enormous, and in the end Lecoq

*Joseph B. Kadane is Leonard J. Savage University Professor of Statistics and Social Sciences, Emeritus, Department of Statistics, Carnegie Mellon University, Pittsburgh, Pennsylvania 15223, USA e-mail: kadane@stat.cmu.edu.*

goes to a wise-man consulting detective who essentially solves the case for him.

Little is written about Lecoq's methods, except for one refrain that occurs three times: "Always suspect that which seems probable; and begin by believing that which appears incredible" (page 79), "Distrust all circumstances that seem to favor your secret wishes" (page 87), and "Always distrust what seems probable!" (page 248). Taken together they suggest a tinge of paranoia, perhaps (but what is a poor detective to do in the hands of a malign author intent on surprising the reader and, one supposes, his own detective?). It also suggests a touch of Lindley's Cromwell's Rule, not to put zero probability on any conceivable possibility (Lindley, 1985, page 104).

We can't go further with Lecoq's maxims or theories, because Gaboriau doesn't give us any.

## 3. EDGAR ALLAN POE'S C. AUGUSTE DUPIN

In *The Murders in the Rue Morgue* Poe (1944), Dupin and a friend read newspaper accounts of two murders in a fourth story room locked from the inside. The mother's throat was slashed many times; the daughter was suffocated and her body stuffed up a chimney. Dupin offers to help the police, and is given access to the crime scene. He finds hair that he is sure is not human. (I won't leave you hanging too long, but first want to introduce you to the problems in the other two Dupin stories.)

In *The Mystery of Marie Rogêt*, a young woman's body is found floating in the Seine River. Using newspaper accounts, Dupin challenges much of the rationale in those stories. The case and the newspaper accounts all did occur in New York City, to a young woman named Mary Cecilia Rogers. Her case is still regarded as unsolved, although there are hints that her death may have been the result of an unsuccessful abortion.

Finally, *The Purloined Letter* is the most famous of the three Dupin stories. A police prefect asks Dupin's help in finding and returning a letter concerning a high-placed lady who saw the letter being taken by Minister D., but was helpless to prevent the theft. Minister D. has since been using the letter for blackmail. The prefect states that the letter has not been revealed, because the consequences that would have ensued from its release have not occurred. Second, Minister D. must have the letter close at hand for it to be useful to him. Finally, using various subterfuges, the police have carefully searched all the hiding places in Minister D.'s rooms, behind the wallpaper, hidden in a hollow leg of furniture, etc., without result. They have also twice found ways to search Minister D.'s body, again without finding the letter.

What is important to us in these three stories is the theory Poe promulgates as to how Dupin is thinking about the puzzles he sets himself to solve. In a preamble to *The Murders in the Rue Morgue*, Poe writes of his views on skill in games. The first of these is chess, which he regards as principally a matter of attention (the loser, by inattention, makes a blunder). (I don't think Poe is correct about chess among decent players, as winners are often those who employ sound openings, develop their pieces, pay attention to their pawn structure, protect their king, fight for control of important center squares, and gradually accumulate little advantages all the while thwarting his opponent's attempts to do the same to him.) Draughts (now called checkers) interests Poe more. For example, if the game is reduced to four kings, he writes "Deprived of ordinary resources, the analyst throws himself into the spirit of his opponent, identifies himself therewith, and not unfrequently sees thus, at a glance, the sole methods (sometimes absurdly simple ones) by which he may seduce into error or hurry into miscalculation" (page 47).

The third game to interest Poe is whist, which is roughly like bridge without bidding.

> ... proficiency in whist implies capacity for success in all these more important undertakings where mind struggles with mind. When I say proficiency, I mean that perfection in the game which includes a comprehension of *all* the sources whence legitimate advantage may be derived. These are not only manifold, but multiform, and lie frequently among recesses of thought altogether inaccessible to the ordinary understanding.
>
> ... it is in matters beyond the limits of mere rule that the skill of the analyst is evinced. He makes, in silence, a host of observations and inferences. So, perhaps, do his companions; and the difference in the extent of the information obtained, lies not so much in the validity of the inference as in the quality of the observation. The necessary knowledge is that of *what* to observe. Our player confines himself not at all; nor, because the game is the object, does he reject deductions from things external to the game. He examines the countenance of his partner, comparing it carefully with that of each of his opponents. He considers the mode of assorting the cards in each hand; often counting trump by trump, and honor by honor, through the glances bestowed by their holds on each. He notes every variation of face as the lay progresses, gathering a fund of thought from the differences in the expression of certainty, of surprise, of triumph, of chagrin. From the manner of gathering up a trick he judges whether the person taking it, can make another in the suit.

The first two or three rounds having been played, he is in full possession of the contents of each hand, and thenceforward puts down his cards with as absolute a precision of purpose as if the rest of the party had turned outward the faces of their own. The analytical power would not be confounded with simple ingenuity; for while the analyst is necessarily ingenious, the ingenious man is often remarkably incapable of analysis.

Between ingenuity and the analytic ability there exists a difference far greater, indeed, than that between the fancy and the imagination, but of a character very strictly analogous. It will be found, in fact, that the ingenious are always fanciful, and the *truly* imaginative never otherwise than analytic (pages 48, 49).

It is this talent that Dupin is called upon to exemplify. For example, how did the murderer or murderers escape from the murder room in *The Murders in the Rue Morgue*? "I knew that all apparent impossibilities *must* be proved not to be such in reality. I proceeded to think, thus—a posteriori. The murderers *did* escape from one of these windows. This being so, they could not have re-fastened the sashes from the inside, as they were found to be fastened... Yet the sashes *were* fastened. They *must*, then, have the power of fastening themselves" (pages 72, 73, emphasis in original). Dupin then goes on to show that while the two windows look the same, one of them could fasten itself. He thus establishes how the murderers escaped.

Putting together the wild brutality of the murders, the enormous strength it must have taken to put the dead daughter's body up the chimney, and the nonhuman hair he found on the scene, Dupin concludes that some ape-like animal must have done these murders. He places an ad asking if someone has lost an "Ourang-Outang." When a sailor shows up, Dupin learns that he had an orangutan which escaped with the sailor's shaving razor, murdered the two women, and escaped.

What can we say of how well Dupin's deductions used Poe's theories? Certainly his deduction about egress is using a combination of Bayes' rule and Lindley's "Cromwell's Rule." However, perhaps we can excuse this omission of the use of his ideas about game theory in the light of the bouquet Poe throws to all the readers of this journal:

> Coincidences, in general, are great stumbling-blocks in the way of that class of thinkers who have been educated to know nothing of the theory of probabilities— that theory to which the most glorious of human research are indebted for the most glorious of illustration (page 77).

As already mentioned, *The Mystery of Marie Rogêt*, remains a mystery. Solving real cases is no doubt more demanding than solving fictional ones. Nonetheless, there is one passage that demands our attention. Poe writes,

> [The journal *L'Etoile* writes] 'All experience has shown that drowned bodies, or bodies thrown into the water immediately after death by violence, require from six to ten days for sufficient decomposition to take place to bring them to the top of the water.'
>
> These assertions have been tacitly received by every paper in Paris, with the exception of *Le Moniteur*.[1] This latter print endeavors to combat that portion of the paragraph which has reference to 'drowned bodies' only, by citing some five or six instances in which the bodies of individuals known to be drowned were found floating after the lapse of less time than is insisted upon by *L'Etoile*. But there is something excessively unphilosophical in the attempt, on the part of *Le Moniteur*, to rebut the general assertion of *L'Etoile*, by a citation of particular instances militating against that assertion. Had it been possible to adduce fifty instead of five examples of bodies found floating at the end of two or three days, these fifty examples could still have been properly regarded only as exceptions to *L'Etoile's* rule, until such time as the rule itself should be confuted. Admitting the rule (and this *Le Moniteur* does not deny, insisting merely upon its exceptions), the argument of *L'Etoile* is suffered to remain in full force; for this argument does not pretend to involve more than a question of the *probability* of the body having risen to the surface in less than three days; and this probability will be in favor of *L'Etoile's* position until the instances so childishly adduced shall be sufficient in number to establish an antagonistical rule (pages 112, 113).

---

[1] The New York *Commercial Advertiser*, edited by Col. Stone.

This is an important and subtle point, one that it took the medical profession another century to incorporate, via the use of controlled clinical trials.

Finally, we come to Poe's masterpiece, *The Purloined Letter*. Dupin further expands on the theory he is using in solving the case by discussing yet another game, as follows:

I knew one [school-boy] about eight years of age, whose success at guessing in the game of 'even and odd' attracted universal admiration. This game is simple, and is played with marbles. One player holds in his hand a number of these toys and demands of another whether that number is even or odd. If the guess is right, the guesser wins one; if wrong, he loses one. The boy to whom I allude won all the marbles of the school. Of course he had some principle of guessing; and this lay in mere observation and admeasurement of the astuteness of his opponents. For example, an arrant simpleton is his opponent, and, holding up his closed hand, asks, 'Are they even or odd?' Our schoolboy replies, 'Odd,' and loses; but upon the second trial he wins, for he then says to himself: "The simpleton had them even upon the first trial, and his amount of cunning is just sufficient to make him have them odd upon the second; I will therefore guess odd';—he guesses odd, and wins. Now, with a simpleton a degree above the first, he would have reasoned thus: 'This fellow finds that in the first instance I guessed odd, and, in the second, he will propose to himself, upon the first impulse, a simple variation from even to odd, as did the first simpleton; but then a second thought will suggest that this is too simple a variation, and finally he will decide upon putting it even as before. I will therefore guess even';—he guesses even, and wins. Now this mode of reasoning in the school-boy, whom his fellows termed 'lucky,'—what, in its last analysis, is it?

'It is merely,' I said, 'an identification of the reasoner's intellect with that of his opponent' (pages 165, 166).

If we identify utilities with marbles, this is a zero-sum two person game. The minimax strategy, independently one-half probability on odds and half on evens, seems a very poor recommendation to this young genius.

Dupin reasons as follows: Minister D. knows the methods of the police. They are extremely good at ferreting out and exploring all of the hidden places the letter might be. None of those searchers has succeeded. Thus Minister D., knowing what he does about the police, does not put it in any of these places. Where then? It must be in plain sight!

To test his idea, Dupin visits Minister D., wearing eyeglasses that obscure where his eyes are focusing. He sees a scruffy document hanging, observes that it is folded inside out, and deduces that this must be the letter. Leaving a gold snuff box so he has an excuse to return the next day, Dupin has a document prepared that matches the new exterior of the letter. Returning the next day, purportedly to pick up his snuff box, a cannon goes off in the street below (which Dupin had arranged). Minister D. is distracted, Dupin switches his fake for the real letter, and leaves with both the real letter and his gold snuff box.

In this story, Poe has finally delivered on his promise. Dupin has used his understanding of Minister D.'s thought process to identify where the letter is. *The Purloined Letter* is a wonderful story. It enchanted me as a child, and still does.

## 4. SIR ARTHUR CONAN DOYLE'S SHERLOCK HOLMES

We do not need to speculate about the antecedents Doyle had in mind. In his autobiography, *Memories and Adventures* Doyle (1924) he writes:

> Gaboriau had rather attracted me by the neat dovetailing of his plots, and Poe's masterful detective, M. Dupin, had from boyhood been one of my heroes. But could I bring an addition of my own? I thought of my old teacher Joe Bell, ... of his eerie trick of spotting details. If he were a detective, he would surely reduce this fascinating but unorganized business to something nearer an exact science (page 69).

In contrast to Gaboriau's single (or perhaps double) book and Poe's three short stories, Doyle (1981) gives us four Sherlock Holmes novels and 56 short stories. So we have in one sense a great deal of information. However, Doyle seems less anxious than Poe to show us how Holmes is thinking about his tasks. When he does so, on occasion those thoughts are often reminiscent of ideas already in Poe's stories. For example, in *The Adventure of the Second Stain* from *The Return of Sherlock Holmes*, a letter from a foreign power has been stolen. If its content were known, it could cause various foreign upsets. "Only one important thing has happened in the last three days, and that is that nothing has happened" (page 659). It seems to me that this is much like the evidence in *The Purloined Letter* that the letter had not been used.

Similarly, in *The Hound of the Baskervilles*, Holmes says "If ... we are dealing with forces outside the ordinary laws of Nature, there is an end to our investigation. But we are bound to exhaust all other hypotheses before falling back upon this one" (page 684). I find this reasoning similar to that Dupin used in ascertaining how the murderer(s) left the scene of the murders in *Murders in the Rue Morgue*.

There is one passage, in the conclusion to *A Study in Scarlet*, where I think Doyle is aiming to clarify what Poe is trying to get at on the subject of analytic reasoning: "Most people, if you describe a train of events to them, will tell you what the result would be. They can put those events together in their minds, and argue from them that something will come to pass. There are few people, however, who, if you told them a result, would be able to evolve from their own inner consciousness what the steps were that led to that result. This power is what I mean when I talk of reasoning backward, or analytically" (pages 83, 84). So Doyle (Holmes) is saying that predicting subsequent from preceding events is relatively straightforward, but the reverse is hard. And this is exactly what Bayes' Theorem does.

However, that theorem is even more evident in what we must take as Holmes' slogan, as it is repeated four times in the work (pages 111, 315, 926, 1011). "When you have eliminated the impossible, whatever remains, however unlikely, must be the truth."

This is formulated sufficiently crisply that it can actually be proved as follows: Let $H_1, \ldots, H_k$ be $k$ theories of the case, mutually exclusive (not more than one can be true), exhaustive (one of them must be true), and suppose each has positive prior probability. In fact, we'll think of $H_1$ as the theory (however unlikely) that is not eliminated by the data. Suppose we have data $X$ that has eliminated theories $H_2, \ldots, H_k$, that is,

$$(1) \quad P\{X|H_i\} = 0, \quad i = 2, \ldots, k,$$

but $P\{X|H_1\} \neq 0$.
Then

$$P\{H_1|X\}$$
$$= \frac{P\{X|H_1\}P\{H_1\}}{\sum_{i=1}^{k} P\{X|H_i\}P\{H_i\}} \quad \text{(this is Bayes' Theorem)}$$
$$= \frac{P\{X|H_1\}P\{H_1\}}{P\{X|H_1\}P\{H_1\}} \quad \text{substituting (1)}$$
$$= 1.$$

Thus no matter how small $P\{H_1\}$ may have been, the posterior probability of $H_1$ under these circumstances is one.

Thus Sherlock Holmes is using, and insisting upon, Bayesian results to explain his actions.

## 5. CONCLUSION

To make this conclusion maximally embarrassing (and hence possibly entertaining to the reader), I present it as if I were being cross-examined by C:

C: Dr. Kadane, is it correct that you are familiar with Bayes' Theorem?

K: Yes.

C: How would you describe Sherlock Holmes' use of Bayesian ideas?

K: Holmes certainly seems to understand the ideas, and how to use them.

C: Is it known whether Doyle had an acquaintance with mathematics sufficient that he might be familiar with a mathematical version of the theorem?

K: It is known that Doyle was trained and qualified as a physician. I do not know the extent to which some math may have been part of that training. Therefore my answer to your question is "I don't know."

C: Very well, whether or not Doyle had that mathematical training, does Holmes, in Doyle's hands, correctly use Bayes' Theorem?

K: Yes, I think he does.

C: Does he make any errors he might have avoided had Doyle had a more mathematical grasp of Bayes' Theorem?

K: None that I have found.

C: Very well, now let's turn to Poe's work. Dr. Kadane, do I understand correctly that you have written a paper about the connection between games and Bayesian theory?

K: Yes, Pat Larkey and I wrote a paper entitled "Subjective Probability and the Theory of Games" (Kadane and Larkey, 1982).

C: Would you tell us briefly the main argument of that paper?

K: Surely. The idea is that if I am playing a game against you, my main source of uncertainty is what you will do. As a Bayesian I have probabilities on what you will do, and can use them to calculate my maximum expected utility choice, which is what I should choose.

C: Is this consistent with what Poe writes about games?

K: Very much so. The marble king of Poe's acquaintance is very good at guessing his opponent's strategy, which is how he winds up with all the marbles in his school. Dupin is successful at understanding Minister D.'s strategy, and hence in finding and retrieving the letter.

C: Is there anything that Poe writes about games that is inconsistent with your theory?

K: No.

C: On the other hand, is there anything in your paper that would have helped Poe had it been available more than a century before it was?

K: Nothing I can think of.

C: Now, is it also the case that you have written on the subject of skill in games, is that correct?

K: Yes. Four of us, the others were Pat Larkey again, Robert Austin and Shmuel Zamir, wrote a paper by that title, published in *Operations Research* (Larkey et al., 1997).

C: Again, briefly, what is this paper about?

K: We create a simplified version of poker, and simulate contests among various strategies for playing the game. One interesting finding was non-transitivity: under certain circumstances, there could be strategies A, B and C, where A is effective against B, B against C and C against A. So there isn't among these, a "best" strategy at all.

C: Is there anything in this paper that would have deepened Poe's understanding of skill in games?

K: I don't think so. I think Poe understood skill in games very well, both in how Dupin outwits Minister D., and in his general introduction. As I explained earlier, I disagree with him about chess, but as a general matter, his view of skill in games is very similar to the one in our papers.

C: So then is it your thought that you have very little news for either Doyle or Poe?

K: Yes, I think that is fair.

C: Then what has been going on in this field for the last 100 or 150 years? Have we gotten nowhere?

K: I don't think that is a fair characterization. What is new is that through the work of Ramsey (1926), de Finetti (1970, 1975), Savage (1954), DeGroot (1970) and Lindley (1985), we now have a general theory of what it means to make good decisions in the face of uncertainty. That theory rests on a few simple principles:

– all sources of uncertainty are modeled probabilistically,
– as data became available, the probability models are updated by conditioning on the observed data,
– when it is required that decisions be made, the optimal decision maximizes expected utility, where the expectation is taken with respect to the current (updated) opinion of the decision maker.

Thus we now understand both Bayes' Theorem and the Bayesian approach to games as special cases of this very general theory. That's what's new.

C: Thank you, Dr. Kadane.

Both detective stories and Bayesian analysis have flourished in the intervening century. They share some common roots.